\documentclass[]{spie}  
\usepackage[]{graphicx}
\usepackage{siunitx}
\usepackage{caption}	
\usepackage{subcaption}	
\usepackage{sidecap}
\usepackage{wrapfig}
\usepackage{hyperref}

\DeclareSIUnit{\cmps}{\cm\per\second}
\DeclareSIUnit{\mps}{\meter\per\second}
\DeclareSIUnit{\kmps}{\kilo\meter\per\second}
\DeclareSIUnit{\micron}{$\mu$\meter}
\DeclareSIUnit{\foot}{'}
\DeclareSIUnit{\inch}{"}
\title{A microlens-array based pupil slicer and double scrambler for MAROON-X} 

\author{Andreas Seifahrt\supit{a}, Julian St\"urmer\supit{a} and Jacob L. Bean\supit{a}
\skiplinehalf
\supit{a}University of Chicago, USA \\
}

\authorinfo{Further author information: (Send correspondence to A.S.)\\A.S.: E-mail: seifahrt@uchicago.edu, Telephone: 1 773 702 9877}

\begin{document} 
  \maketitle 

\begin{abstract}
We report on the design and construction of a microlens-array (MLA)-based pupil slicer and double scrambler for MAROON-X, a new fiber-fed, red-optical, high-precision radial-velocity spectrograph for one of the twin 6.5m Magellan Telescopes in Chile. We have constructed a 3X slicer based on a single cylindrical MLA and show that geometric efficiencies of $\geq85\%$ can be achieved, limited by the fill factor and optical surface quality of the MLA. We present here the final design of the 3x pupil slicer and double scrambler for MAROON-X, based on a dual MLA design with (a)spherical lenslets. We also discuss the techniques used to create a pseudo-slit of rectangular core fibers with low FRD levels.

\end{abstract}



\section{INTRODUCTION}

MAROON-X is a new fiber-fed, red-optical, high-precision radial-velocity spectrograph for one of the twin 6.5m Magellan Telescopes in Chile, currently under construction at the University of Chicago\cite{seifahrt}. MAROON-X is based on a KiwiSpec R4-100 echelle spectrograph\cite{barnes}, which has a \SI{100}{\milli\meter} beam diameter that yields a resolution-slit product of $R\phi\approx25,400$'' for a 6.5m telescope. In order to achieve the desired resolving power of 80,000 and an acceptable field of view on sky, we either needed to slice the image or the pupil. Either technique will boost the efficiency of the spectrograph at the cost of an increase in slit height and thus reduced spectral coverage and increased aberrations for a given detector size.

We decided against image slicing, since acceptable efficiencies are often hard to achieve and the on-sky performance of image slicers remains practically untested in the context of high-precision radial velocity work. It is important to realize that an image slicer effectively works as an anti-scrambler, as it non-linearly amplifies small changes in the fiber output illumination at the slicer edges. Moreover, being directly in the imaging plane of the spectrograph, a sub-m\,s$^{-1}$ stability requirement translates into nanometer levels of positional stability, which is challenging to control both on a mechanical and thermal level.

We thus decided to built a pupil slicer, which is much less critical in terms of (thermo-)mechanical stability. This technique has been selected for the next generation of mid-size (e.g. SPIRou/CFHT\cite{spirou}) and large spectrographs (i.e. G-CLEF/GMT\cite{gclef,gclef2} and ESPRESSO/VLT\cite{espresso,espresso2}).

A wide variety of design implementation for pupil slicers are possible. Their complexity is partly driven by the number of slices required to achieve a certain resolution-slit product at the spectrograph. A simple 2x slicer can for example be built using a Bowen-Walraven image slicer\cite{walraven} in the pupil plane and re-imaging the sliced pupil onto a rectangular fiber\cite{spronck}. For geometrical reasons such a 2x slicer already requires fibers with a 1:4 aspect ratio, practically limiting this approach to two slices. Much higher slicing factors typically require designs that exploit anamorphic (de-)magnifications for more complex reformatting of the pupil\cite{espresso}. 

For MAROON-X we require a 3x slicer to meet our resolving power and throughput requirements. We settled on a design first suggested more than 15 years ago: Using microlens arrays (MLAs) in combination with fibers to slice the pupil into $n$ slices and feed them into the same number of fibers which are re-arranged into a slit\cite{iye}. Depending on the number of slices and the desired re-imaging technique, hexagonal (a.k.a. fly's eye), cylindrical, and spherical MLAs can be used.

We started with a prototype design using one cylinder lens and a cylindrical MLA. We show results from lab measurements with this design in Section\,\ref{cylinder}. Encouraged by these results we expanded on the design to include a double scrambler and to improve on the efficiency of the slicer. Our final design and its implementation for MAROON-X are shown in Section\,\ref{final} and efficiency estimates are given in Section\,\ref{efficiency}.

\section{PROTOTYPE PUPIL SLICER}\label{cylinder}

In order to evaluate the suitability of MLAs for a pupil slicer, we started with a simplified design of a 3x slicer based on a single cylinder lens and a cylindrical MLA. In this design, a collimator forms a pupil image at the front face of the cylindrical MLA, covering the full width of three of the lenslets in the array. 
\begin{figure}[b!]
\centering
\includegraphics[width=0.96\linewidth,clip]{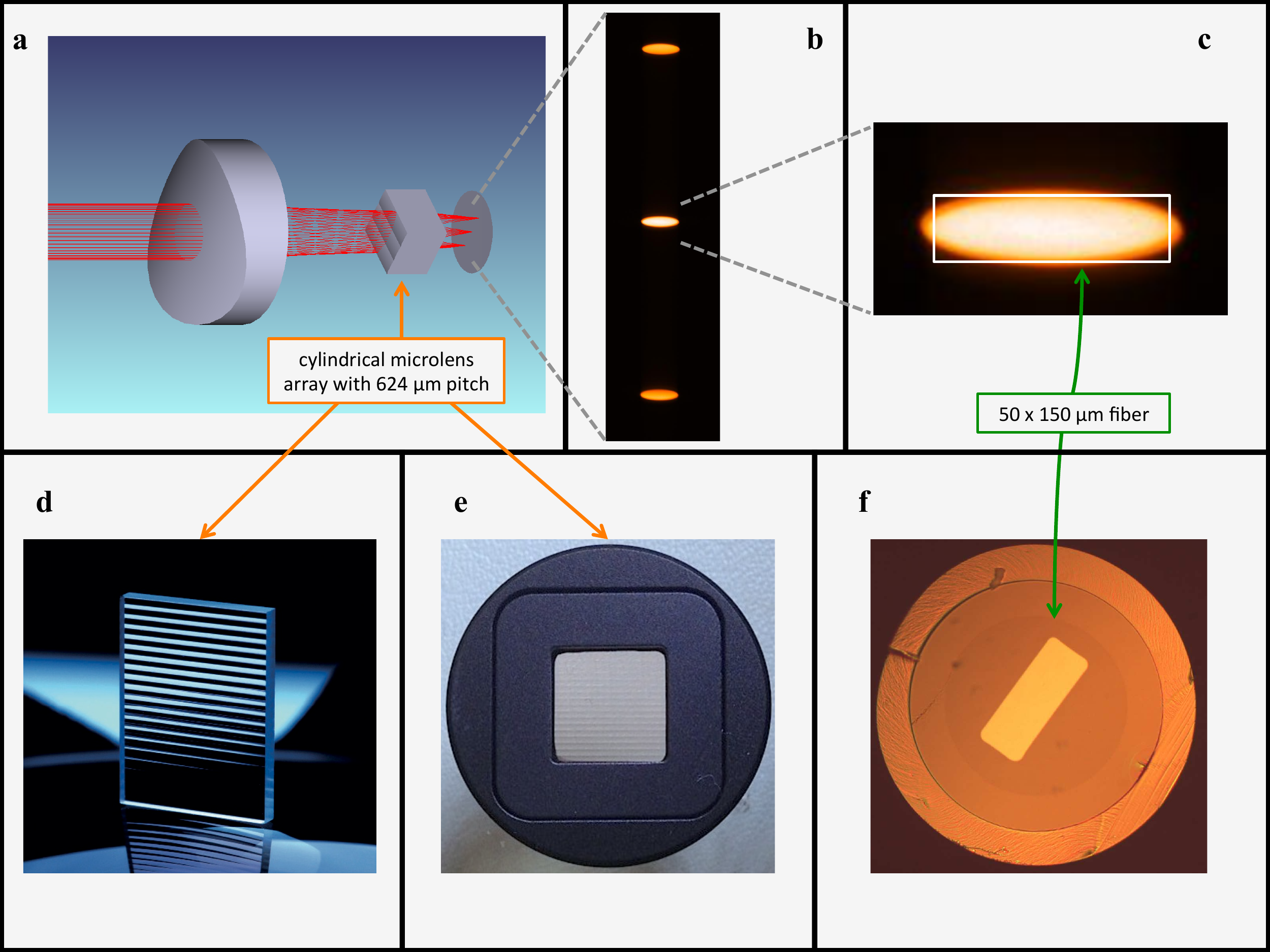}
\vspace{1mm}
\caption{\textbf{Pupil slicer concept and hardware.} \textbf{(a)} Zeemax 3D raytrace of the prototype pupil slicer. A field lens forms a 1.9\,mm diameter image of the telescope pupil (not shown here, off to the left of the image). A single cylindrical lens and a cylinder lenslet array (rotated 90$^{\circ}$ in respect to the single cylindrical lens) picks up the beam. When matching the focal ratios, only three lenslets in the array are illuminated and the rest of the array is not shown for clarity. At the end, three elliptical (anamorphic) images of the star are formed in the image plane at $f/4.66$ to be fed into three separate rectangular fibers (fibers not shown). The output ends of the fibers will be stacked to create a pseudo slit at the spectrograph entrance. \textbf{(b)} Image of the output of the fiber slicer as measured in our lab. The spots are the three elongated images of an evenly illuminated 300\,$\mu$m pinhole, representing a 1" field of view on the sky. \textbf{(c)} Expanded image of the central spot with the footprint of a 50 x 150\,$\mu$m fiber overlayed (white line). \textbf{(d)} Example microlens array (Image courtesy of \textit{INGENERIC GmbH}). \textbf{(e)} Actual microlens array used in our feasibility study. \textbf{(f)} Image of an example rectangular fiber that will be used for the system (Image courtesy \textit{j-fiber GmbH}).}
\label{slicer}
\end{figure}
\clearpage
The cylinder lenslets re-focus the light in one dimension behind the MLA. A single cylinder lens with three times the effective focal length (EFL) of the MLA lenslets, placed before the MLA and rotated by \ang{90}, focuses the light in the perpendicular direction onto the same focal plane. The resulting cone angles are identical in both directions. A circular fiber or pinhole is thus re-imaged into three elliptical images with 1:3 aspect ratio and a separation equal to the pitch of the lenslets in the MLA (see Figure\,\ref{slicer}).

Fibers with rectangular cores placed at the position of the images receive the light and can be later re-arranged to a long-slit to feed the light into the spectrograph. One of the constraining factors is thus the cladding size of the fibers, which limits the proximity of the fibers and thus of the individual images, both at the slicer and at the spectrograph.

For the prototype we used stock optics, including three MLAs with varying EFLs and varying pitch (i.e. varying separation of the lenslets) from Advanced Microoptic Systems GmbH (Germany). These MLAs are made from high-index glasses (S-TiH10 and S-TiH53) in a photo-lithographic process. We chose MLA parameters that allowed for an optical design close to our desired input and output \'etendue for MAROON-X. Typical lens pitches ranged from \SIrange{500}{1500}{\micron} with radii of curvature from \SIrange{2.2}{5.1}{\milli\meter}. We found matching single cylinder lenses and Hastings triplet collimators and used a combination of color filter, iris, and pinhole to inject light with known bandpass and $f$-number. At the output image plane we used a 10x Mitutoyo M Plan objective and a SBIG STF-8300 CCD camera to record the resulting images. A careful distribution of the available degrees of freedom among the individual optical elements allowed a precision alignment of the setup.

For one of the MLAs we tested the throughput and alignment tolerances. This MLA came closest to the desired output configuration for MAROON-X, delivering $f/4.66$ (instead of the desired $f/5$) with an image size almost exactly identical to the size of the rectangular fibers intended for MAROON-X (\SI{50x150}{\micron}).
\begin{SCfigure}[][b!]
\centering
\includegraphics[width=0.6\linewidth,clip]{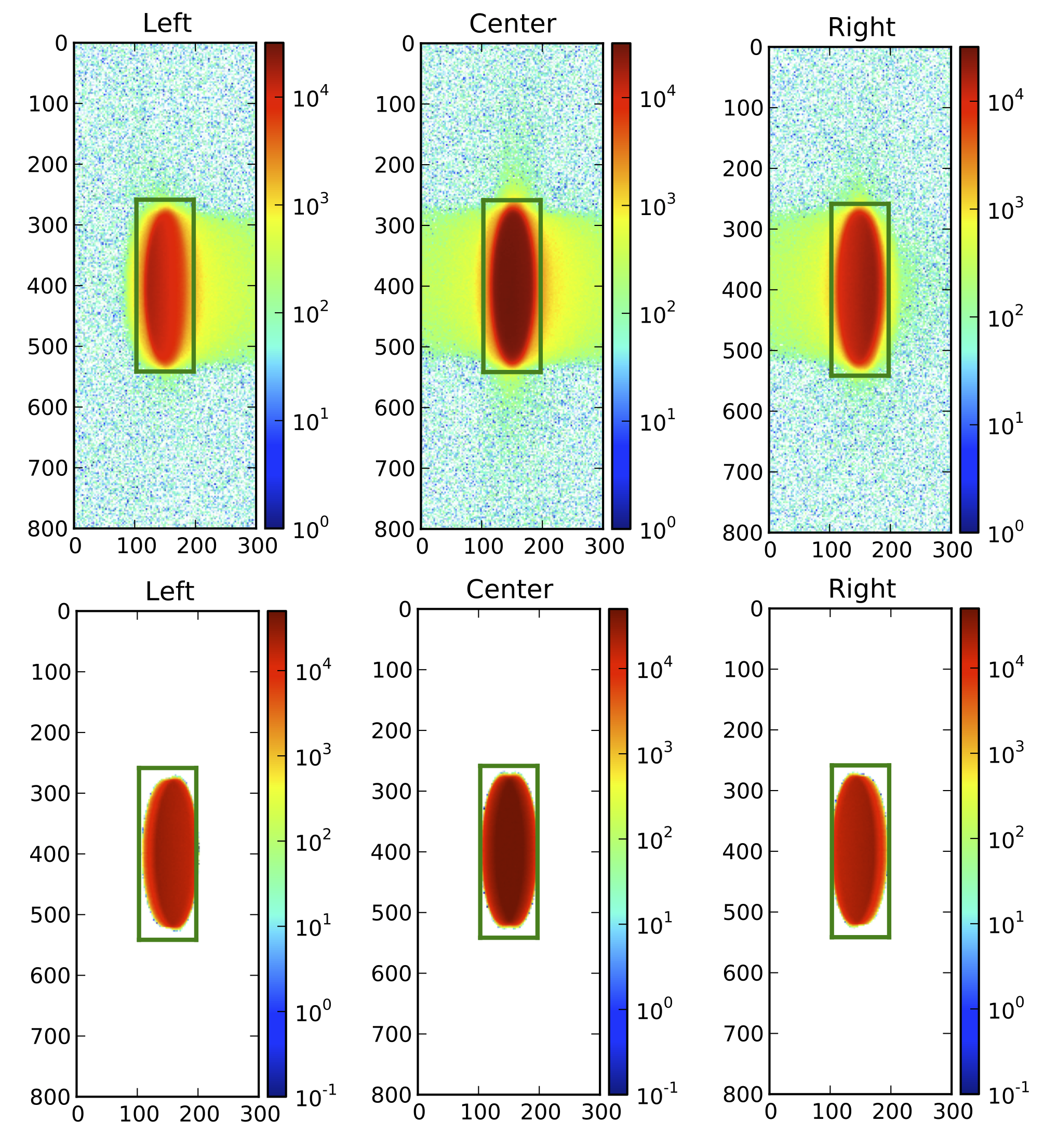}
\caption{\textbf{Images from the prototype pupil slicer} show the flux (in logarithmic scaling) of the three anamorphic images as measured (top row) and from a Zemax image simulation (bottom row). The green boxes represent the aperture of our square fibers (\SI{50x150}{\micron}). The slicer was aligned for 650\,nm and light with a 40\,nm bandpass around 650\,nm was chosen for this particular measurement. The image $f$-number is 4.66. Scattered light between the slices at about 1\% of the peak intensity is visible in the measured images. We find a geometrical efficiency of 86\% for this setup.}
\label{comparison}
\vspace{0mm}
\end{SCfigure}
A full image of the three slices is shown in Figure\,\ref{slicer}(b). We show part of the analysis of the resulting images in Figure\,\ref{comparison} where we measure the flux inside the nominal aperture of our rectangular fibers and compare it with the total flux in the image plane. We compare this to simulated images from Zemax for optimal alignment conditions. 

For a fixed alignment, we find geometrical efficiencies (the fraction of flux collected by the three \SI{50x150}{\micron} vs. the total flux in the image field) of 83--86\%. The variation is due to chromatic effects over our total bandpass of 500--900\,nm. We find that scattered light from the zone between the cylinder lenslets in the MLA is one of the limiting factors in achieving optimal geometrical throughput. The nominal fill factors of the MLAs we tested vary between 91--96\%. Additional losses are thus coming from imperfect alignment and scattering on the lenslet surfaces.

\section{MAROON-X pupil slicer and double scrambler}\label{final}
\subsection{MLA design and production}
Based on the encouraging results obtained with our prototype, we improved upon our design in two ways. First we combined the individual cylinder lens and the MLA into one double-sided MLA to save two air-glass surfaces. We then further modified the design to project the sliced pupil instead of the anamorphic image onto the output fibers.
\begin{figure}[b!]
\centering
\begin{minipage}[t][][t]{0.8\linewidth}
\includegraphics[width=0.5\linewidth,clip]{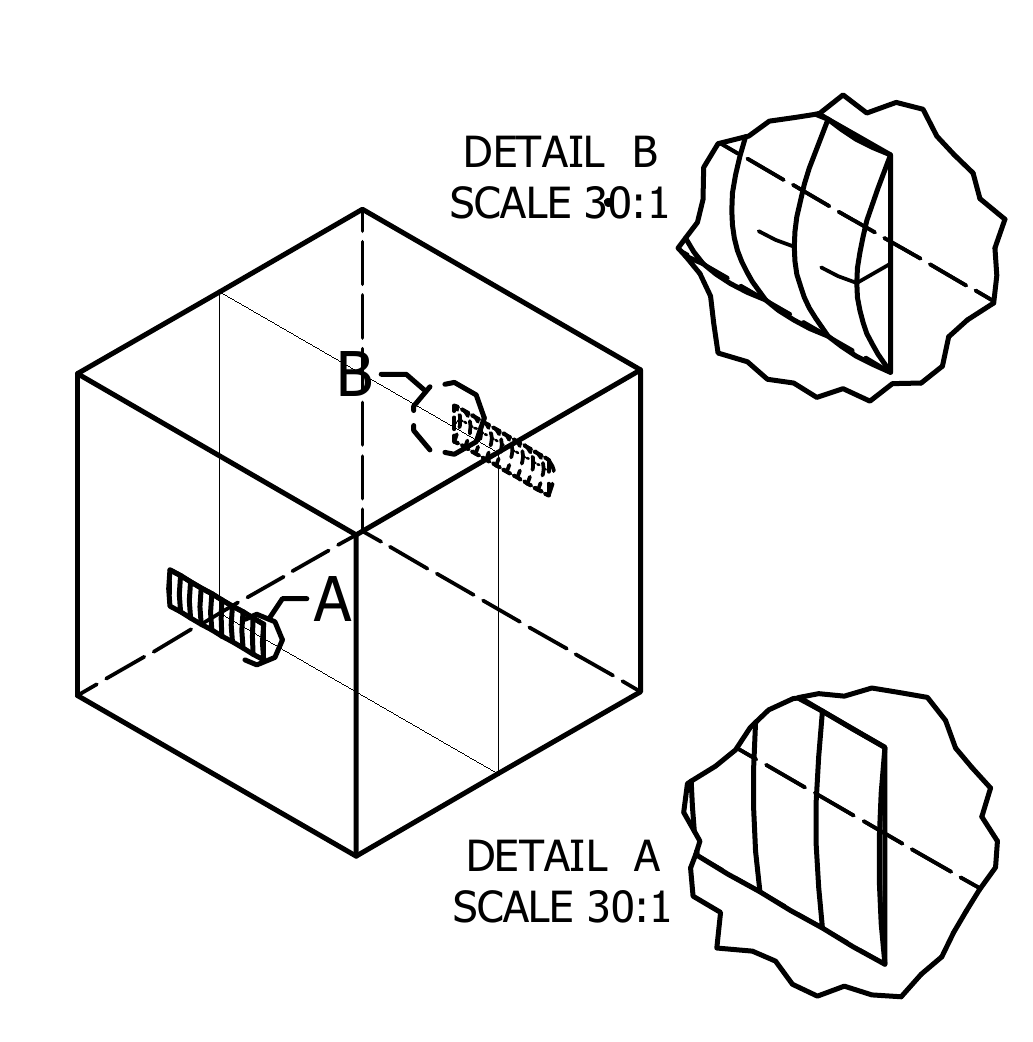}
\includegraphics[width=0.5\linewidth,clip]{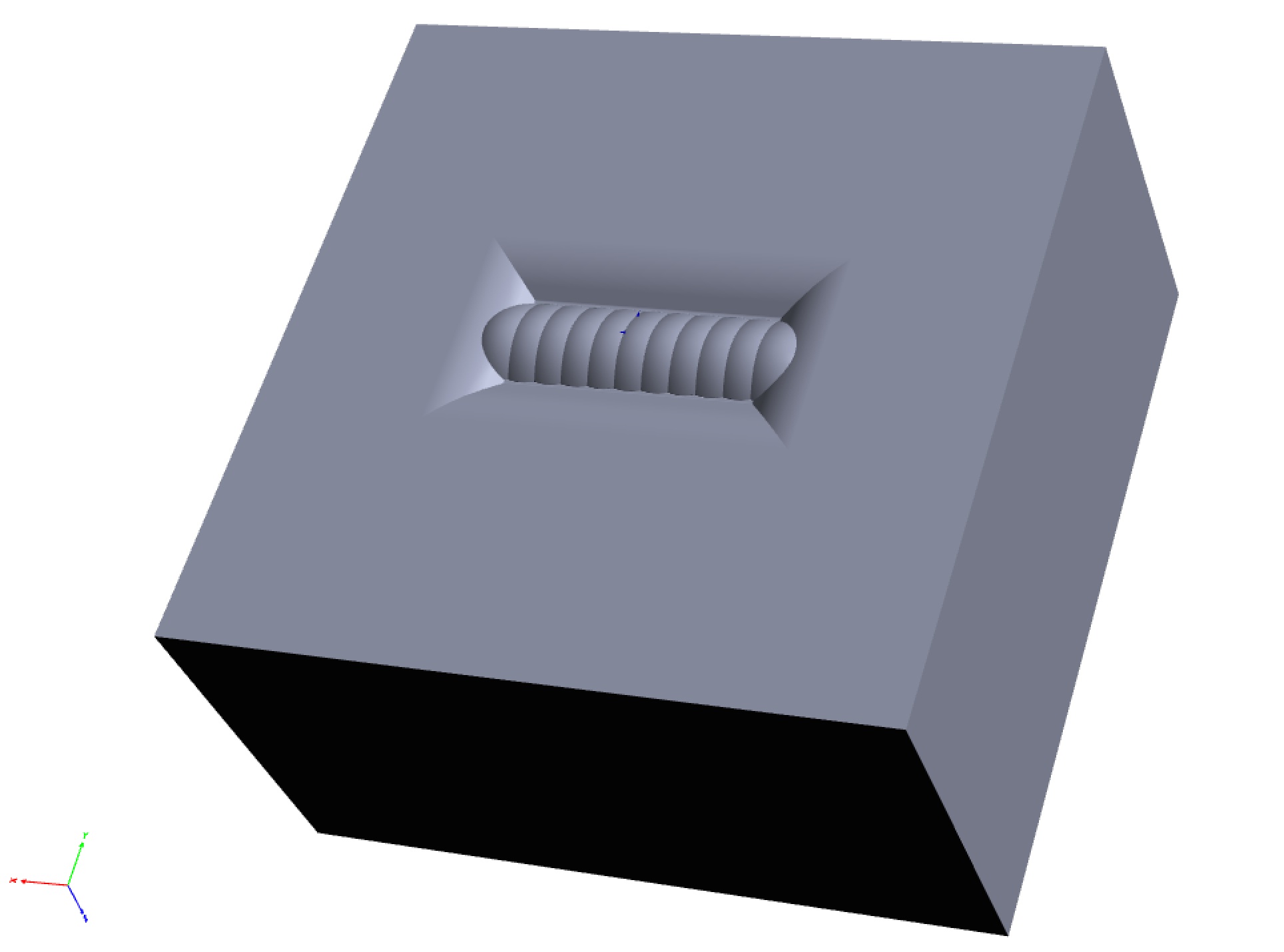}
\end{minipage}

\vspace{2mm}

\includegraphics[width=0.4\linewidth,clip]{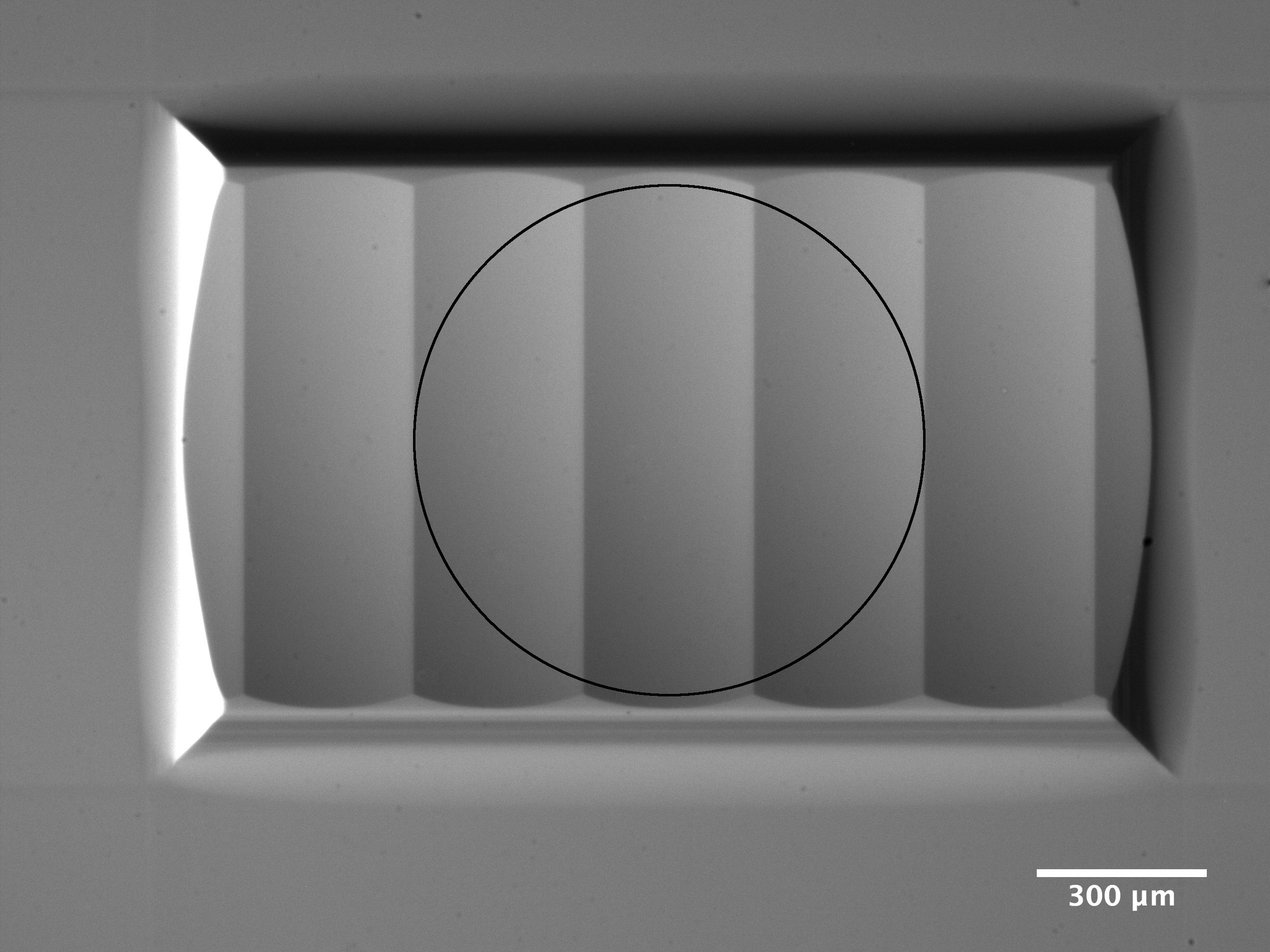}
\includegraphics[width=0.4\linewidth,clip]{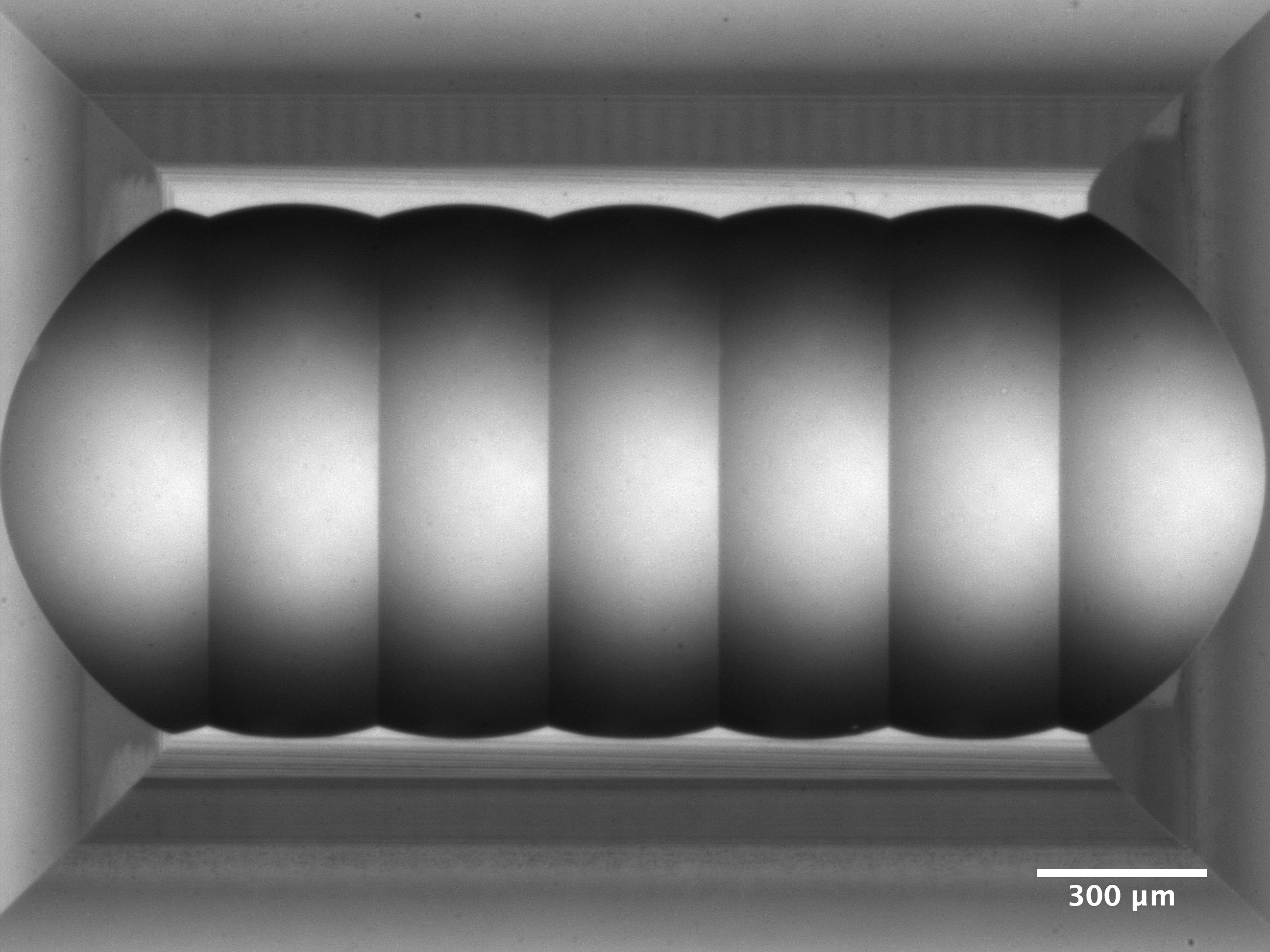}

\vspace{2mm}

\caption{\textbf{Molded MLAs for the MAROON-X pupil slicer.} The MLA chosen for MAROON-X has lenslets on two faces (top left) and is molded in two parts. A pre-production rendering of one of the MLAs is shown in the top right corner. Microscope images of the as-built lenslets under diffuse illumination are shown in the bottom row for the front (left) and back (right) lenslets. Only three of the lenslets of each MLA will be used. The footprint of the pupil on the first MLA is shown in the lower left image as a black circle.}
\label{MLA}
\vspace{0mm}
\end{figure}
By feeding a pupil image into the slit-forming fibers, we effectively incorporate a double scrambler\cite{scrambler} into the pupil slicer as we use the input fiber to scramble the stellar image and the output fibers to scramble the (sliced) pupil image. 
This also benefits the illumination stability of the slicer itself, as temporal instabilities in the slicing geometry will effectively be reduced by the subsequent scrambling of the pupil images.

Our final MLA design is based on two sets of spherical lenslets with \SI{300x900}{\micron} apertures and radii of curvature of \SI{3.3}{\milli\meter} and \SI{0.53}{\milli\meter} for the front and back surfaces, respectively. The lenslets on the back have an aspheric surface with a conic constant of $-1.5$ to improve the aberrations and reduce the surface sag. Due to these specifications and the requirements on improved fill factor ($>95\%$) we decided against photo-lithographically produced MLAs and for molded optics by Ingeneric (Germany). The substrate is again a high-index glass, this time SUMITA K-VC89 ($n_d=1.81$). Due to the high thickness of the MLA (\SI{8}{\milli\meter}), the MLA is produced in two sections and needs to be aligned and bonded after production. Design drawings and microscopic images of the as-built MLAs are shown in Figure\,\ref{MLA}. The MLAs are coated with a BBAR coating with a reflectivity of $R\leq0.5\%$ over 500--900\,nm. The internal transmittance is $T\geq99.5\%$ for a 8\,mm path length over this wavelength range.

\subsection{Collimator and Fiber Array}

The complete pupil slicer and double scrambler consists of an input fiber feed and collimator and a linear array of rectangular output fibers that form a pseudo-slit as input for MAROON-X. The collimator is a custom \SI{2.5}{mm} diameter plano-convex doublet produced by Linos. It is glued onto our input fiber, a \SI{100}{\micron} diameter octagonal fiber from CeramOptec (OCT-WF100/140/250, NA=0.22) and forms a \SI{900}{\micron} diameter collimated beam from the $f/3.33$ cone of light emerging from the fiber. 

The MLA forms three sliced pupil sections at $f/5$ with a separation of \SI{300}{\micron}. These pupil images are projected onto three CeramOptec fibers with \SI{50x150}{\micron} rectangular cores and \SI{300}{\micron} round claddings (WF 50x150/300N), forming a linear array with the short sides of each fiber core lined up. After a run of approximately \SI{1}{\meter}, in which the fibers are rotated by \ang{90}, another linear array is formed, this time with the long sides of the fiber cores lined up to form the physical entrance slit of the spectrograph (see Figure\,\ref{slit}). At this point, two more rectangular fibers are added to the three object fibers at either end of the slit to add sky- and calibration light. 

On both ends of the short rectangular fiber run we need to align the fibers into a linear array. The position of the individual fiber cores is critical, particularly at the pupil slicer, as we have no way to adjust the relative position of the individual pupil images. To keep the geometrical losses to a minimum, relative positional tolerances of a few \si{\micron} have to be achieved. At the same time, the fibers need to be as close as possible to each other, which excludes an articulated mounting solution. 

\begin{SCfigure}[][b!]
\centering
\vspace{0mm}
\includegraphics[trim={0cm 0cm 0cm 0cm},width=0.6\linewidth,clip]{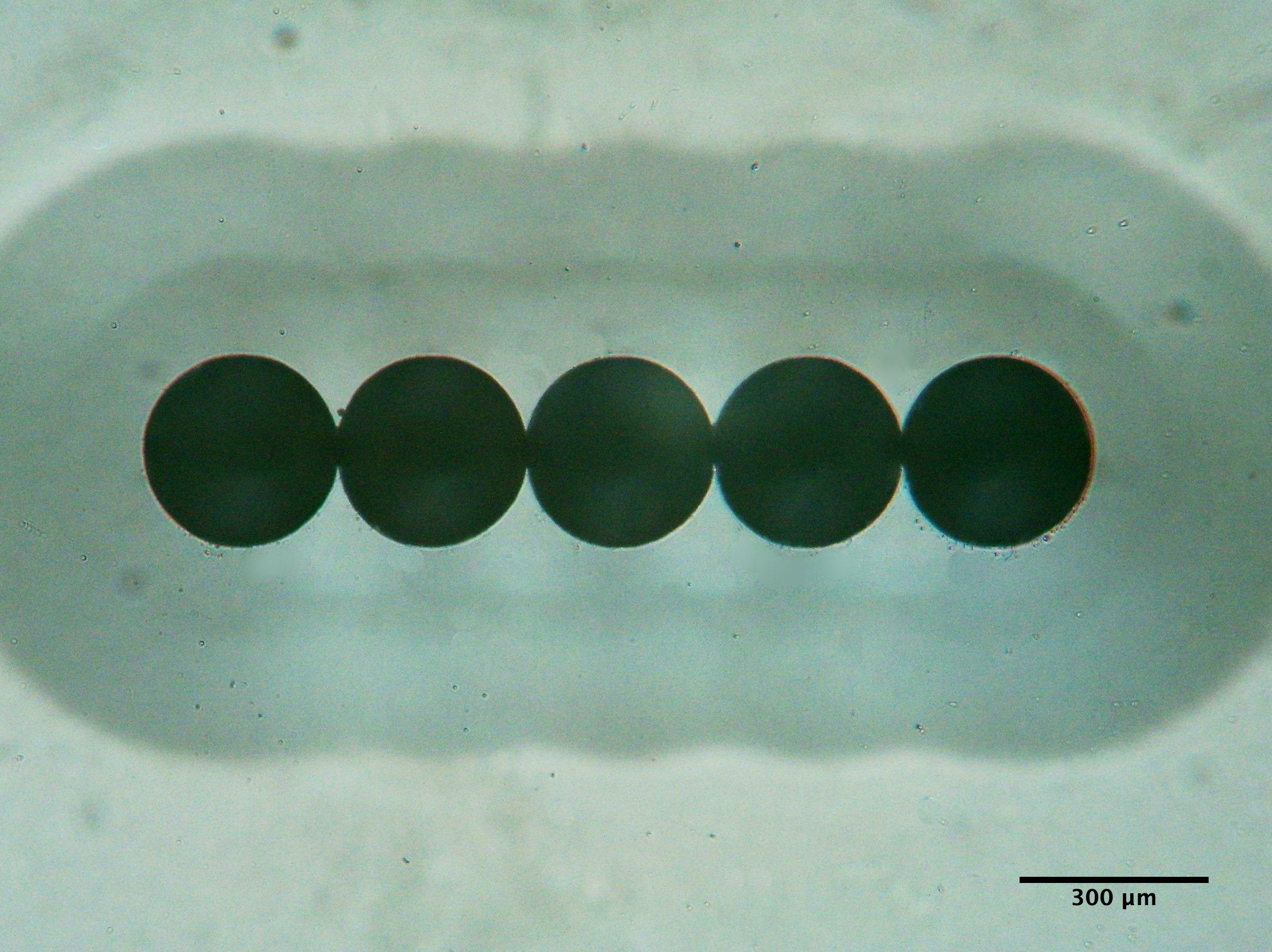}
\vspace{0mm}
\caption{\textbf{Prototype slit plate from FEMTOprint.} The slitplate has \SI{300}{\micron} diameter holes at a \SI{300}{\micron} pitch in a \SI{2}{\milli\meter} fused silica plate. The funnels that allow the insertion of the fibers is visible as the shadow in the background. This version of the slit plate was subsequently replaced with a version having slightly smaller holes at the same pitch (see Figure\,\ref{slit}). 
}
\label{slitplate}
\end{SCfigure} 

A v-groove would be a natural choice for mounting the fibers into a linear array with precisely controlled pitch. This would require the fibers to be either cleaved (which is difficult for fibers with non-circular core structures) or bare-polished. We decided instead for a \SI{2}{mm} thick fused silica plate with precision etched holes produced by FEMTOprint (Switzerland). These slit plates have hole diameter and positional tolerances at the \SI{1}{\micron} level. A funnel at the back-side of the plate helps inserting the bare fibers into the holes. 

Our first prototype for a MAROON-X slit plate had \SI{300}{\micron} diameter holes in a \SI{300}{\micron} pitch, i.e. an overlapping hole pattern (see Figure\,\ref{slitplate}). We found it impossible to insert the fibers because the etches of the overlapping holes were so sharp that they slightly scratched and caught the fibers. Other prototypes with \SI{125}{\micron} hole diameter at the same pitch worked perfectly for standard single- and multi-mode fibers. 
To keep the pitch of the fibers within specification, we decided to etch the fibers in a buffered oxide solution to reduce the cladding diameter from \SI{300}{\micron} to slightly less than \SI{260}{\micron} and comfortably fit the fibers in a slit plate with \SI{260}{\micron} hole diameter (see Figure 5). 

\begin{figure}[!b]
\centering
\begin{minipage}[t]{1\linewidth}
 \includegraphics[trim={0cm 0cm 0cm 0cm},width=0.6\linewidth,clip]{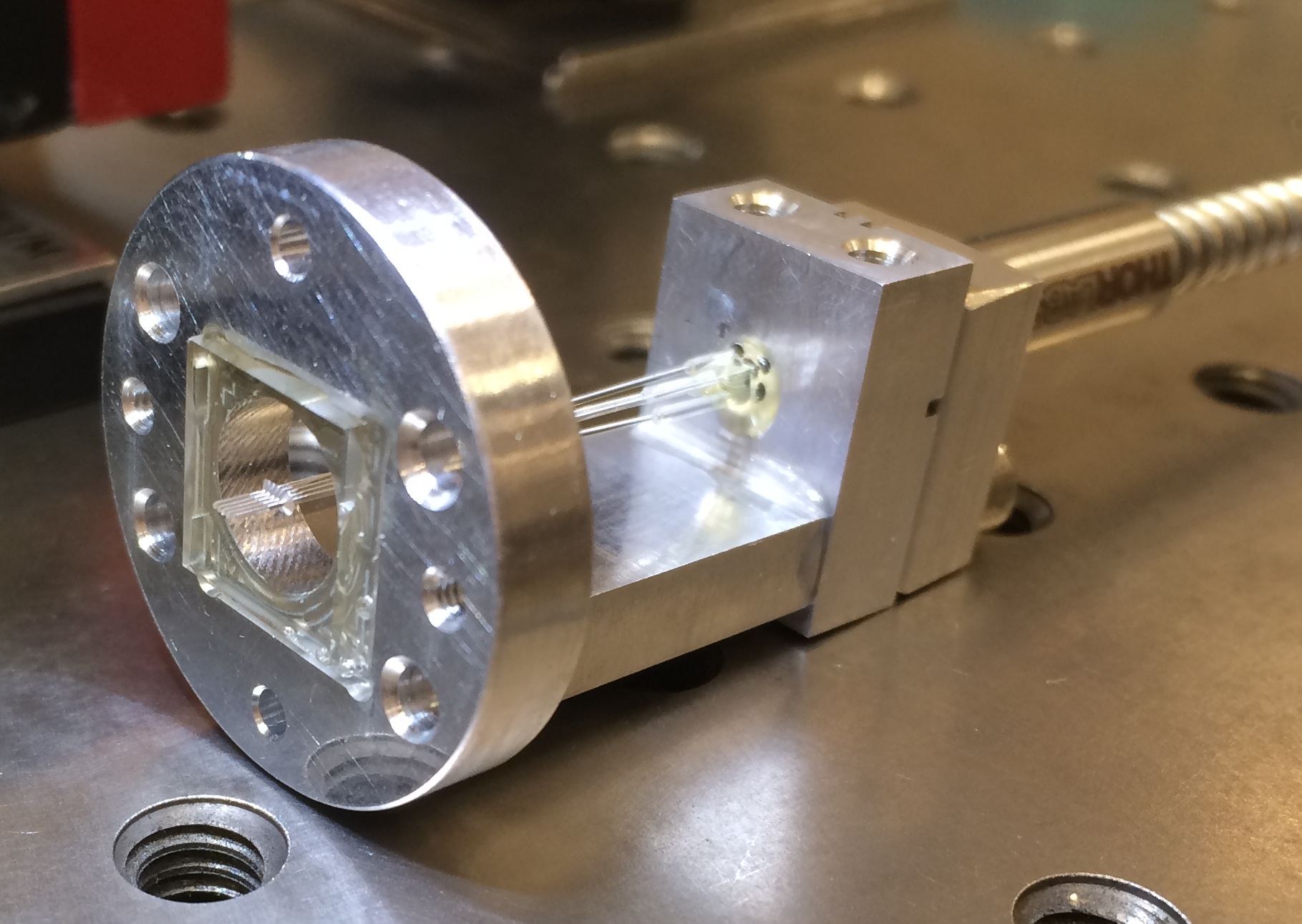}
  \includegraphics[trim={1cm 15cm 3.5cm 0cm},width=0.39\linewidth,clip]{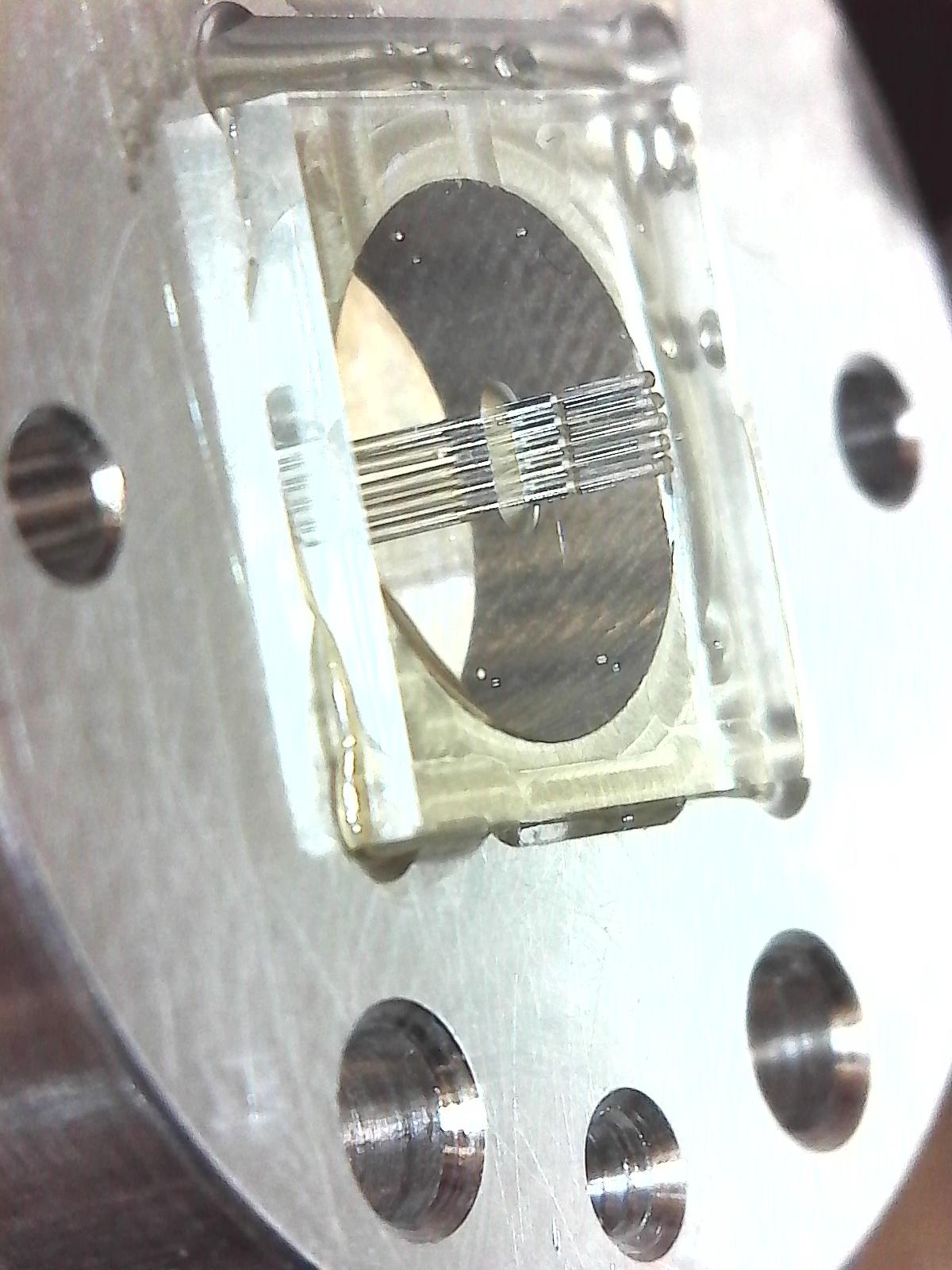}
\end{minipage}

\vspace{1.00mm}

\begin{minipage}[t]{0.6\linewidth}
\hspace{0mm}
 \includegraphics[trim={0cm 0cm 0cm 0cm},width=1\linewidth,clip]{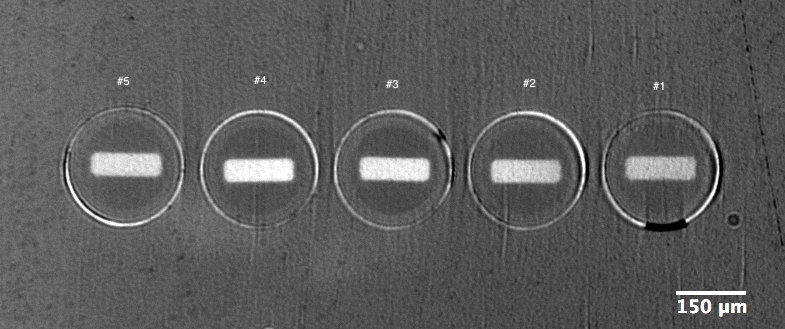}
\end{minipage}
\vspace{2mm}
\caption{\textbf{Prototype linear fiber array and pseudo-slit for MAROON-X}. Top left: FEMTOprint fiber slit plate in a custom mount fixture with five Ceramoptec \SI{50x150}{\micron} rectangular fibers already inserted. Top right: Close-up of the slit plate. The fibers stick out a couple mm from the front of the plate. At this step the fibers are already glued into the guide block for strain relief and to fix their rotation angle but adhesive is not yet applied to the bare fiber ends in the slit plate. Bottom: Same assembly after polishing. While technically within specification, the prototype slit plate shown here has still sub-optimal alignment. The fibers were etched slightly too long, making them \SIrange{5}{6}{\micron} smaller than the holes in the plate, which leads to offsets. Likewise, rotational alignment of two fibers (\#3 and \#4) is off by \SI{-1.5}{\degree} and \SI{+1.1}{\degree}, respectively.
}
\label{slit}
\end{figure}
The fibers are aligned in rotation when being inserted through a guide block and placed in the hole in the slit plate. We back-illuminated each fiber and looked at the output face with a microscope and adjusted the rotation with the fiber clamped in a v-groove on top of a precision goniometer stage. After the rotation is adjusted, each fiber is held in place with a soft UV curing adhesive applied individually to each fiber in the guide block (see Figure\,\ref{slit}a). 

When all fibers are inserted and secured, they are glued into the slit plate with an ultra-low shrinkage room-temperature cure epoxy. We then polish the fibers in the slit plate using a modified Buehler FibrMet polishing machine. An image of a finished slit plate prototype with five of the rectangular fibers forming a pseudo-slit is shown in the bottom of Figure\,\ref{slit}.  

Achievable tolerances for position and rotation of each fiber are \SIrange{2}{3}{\micron} and \SI{\pm0.3}{\degree}, respectively. The final slit plate for MAROON-X will be produced to these specs. Likewise, the polishing quality of the fibers in the prototype slit plate is not quite optimal. This is however tolerable, as a wedge prism is later bonded to the slit plate as part of the MAROON-X spectrograph optics. The adhesive is filling in remaining scratches.

\subsection{Mechanics \& Alignment}

The alignment of the individual components, particularly the bonding of the two MLAs is critical, with tolerances of only \SIrange{2}{3}{\micron}. We plan to use our modified JMAR Mirage 3D microscope stage for this process, which allows lateral positioning with sensitivities of \SI{<1}{\micron} and rotational adjustments with \SI{<15}{''}, respectively. We plan to either adjust and bond the MLAs based on maximized throughput of the complete pupil slicer assembly, or based on laser metrology with a Keyence LT-8110 laser distance measurement system (\SI{2}{\micron} spot size, \SI{0.1}{\micron} resolution).

Since the pupil slicer will be mounted in the vacuum chamber of MAROON-X, all mechanical components need to be vacuum compatible. We plan to use a $XY$ flexure mount for the fiber and collimator assembly, a rotational mount ($\Theta_Z$) for the MLA and a mount with 5 degrees of freedom ($X$,$Y$,$\Theta_X$,$\Theta_Y$, and $Z$) for the output fiber assembly.

\subsection{Efficiency}\label{efficiency}

\begin{SCfigure}[][!b]
\centering
\vspace{0mm}
\includegraphics[trim={1cm 0cm 1cm 0.5cm},width=0.65\linewidth,clip]{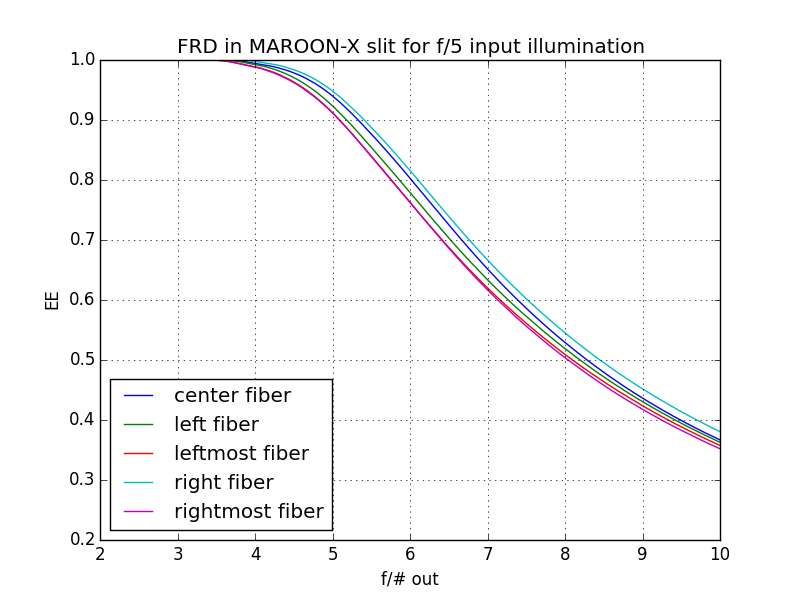}
\vspace{0mm}
\caption{\textbf{FRD measurement for the rectangular fibers in the MAROON-X slit plate.} Fibers are illuminated with white light from a flat input pupil at $f/5$ on one end and their output pupil is imaged after the pseudo-slit formed in the slit plate on the other end. The encircled energy (EE) is measured for varying output pupil apertures for the five fibers in the slit. Fibers with the strongest bending leading up to the slit show the worst FRD. Due to a slight mis-alignment between the slit plate and the guide block, the central fiber is not the fiber with the smallest bending radius. We find EE values of 91\%--96\% for an $f/5$ output aperture.
}
\label{FRD}
\end{SCfigure}

Based on the results of our prototype MLA setup and first indications that the as-built filling factor of our molded MLAs exceeds 98\%, we hope to achieve a geometrical throughput of $\geq90\%$ over the full wavelength range of MAROON-X (500--900\,nm). 

The pupil slicer and double scrambler unit has only four air-glass surfaces, three of which are BBAR coated. The slit plate with the rectangular fibers is uncoated, but a thin fused silica plate with BBAR coating could be bonded on top of the slit plate to reduce Fresnel losses. Specifications on the BBAR coatings on the collimator doublet and the MLAs call for $R\leq0.5\%$ over 500--900\,nm. We thus estimate the efficiency of the scrambler and double slicer to be $\approx$85\%.

Additional losses from focal-ratio-degradation (FRD) effects are hard to quantify in advance, as they depend on a number of factors related mainly to the mechanical stress of the fibers at their mount points and internal stresses. The latter is particularly pronounced for rectangular fibers with rectangular cladding. We hypothesize that the rectangular cladding is putting mechanical stress on the fiber core which results in increased FRD. The fiber we have chosen for MAROON-X (CeramOptec WF 50x150/300N) has a round double cladding and we measure a very low FRD for this fiber\cite{adam}. 

The predominant stress factor is the shrinkage of the adhesive at the fiber mount point, i.e., in the ferrule of a connectorized fiber or the hole in the slit plate. We are using an ultra-low shrinkage adhesive (0.4\% shrinkage) for this reason. 

An additional source of stress in our slit plate assembly is the bending of the bare fibers on the last few mm before entering the slit plate. Since the pitch of the fibers (\SI{300}{\micron}) is smaller than the  coating diameter (\SI{740}{\micron}), the fibers are bent between the guide block and the slit plate (see top of Figure\,\ref{slit}). To characterize this effect, we have measured the FRD of our rectangular fibers with one end glued into the slit plate, the other end in FC/PC connectors. The fiber run was 1\,m, the same as in our final pupil slicer application. We illuminated the fibers with white light at $f/5$ and imaged the output pupil. Details of this setup can be found elsewhere in these proceedings\cite{adam}. The results are shown in Figure\,\ref{FRD}. We find encircled energies (EE) ranging from 91\%--96\% for an $f/5$ output aperture. Since we eventually will have slit plate assemblies on both ends of the rectangular fiber, the combined efficiency for a $f/5$ fiber relay is 88\%--92\% for the central three slices which carry the stellar light from the pupil slicer reducing the overall efficiency of the slicer and double scrambler unit to $\approx$77\%.

\acknowledgments     

We acknowledge funding for this project from the David and Lucile Packard Foundation through a fellowship to J.L.B.


\bibliography{report}   
\bibliographystyle{spiebib}   

\end{document}